# Can observations inside the Solar System reveal the gravitational properties of the quantum vacuum?

Dragan Slavkov Hajdukovic[1]
PH Division CERN
CH-1211 Geneva 23
dragan.hajdukovic@cern.ch
[1]On leave from Cetinje, Montenegro

Abstract
The understanding of the gravitational properties of the quantum vacuum might be the next scientific revolution. It was recently proposed that the quantum vacuum contains the virtual gravitational dipoles; we argue that this hypothesis might be tested within the Solar System. The key point is that the quantum vacuum ("enriched" with the gravitational dipoles) induces a retrograde precession of the perihelion. It is obvious that this phenomenon might eventually be revealed by more accurate studies of orbits of planets and orbits of the artificial Earth satellites. However, we suggest that potentially the best "laboratory" for the study of the gravitational properties of the quantum vacuum is the recently discovered dwarf planet Eris with its satellite named Dysnomia; the distance of nearly *100AU* from the Sun makes it the unique system in which the precession of the perihelion of Dysnomia (around Eris) is strongly dominated by the quantum vacuum.

In physics, the study of particles and atoms has been essential for the understanding of the macroscopic world. In a similar way, the observations within the Solar System (as a miniscule part of the Universe) have been crucial in the understanding of gravitation (including both the Newton law and General Relativity) and hence in the understanding of the Universe. However, it is history of physics and nearly no one believes that the present day observations of the Solar System have potential to reveal the secrets of the Universe at much larger scales like galaxies and clusters of galaxies. I wonder if it is so, or there are hidden potentials and surprises. Can the Solar System, play again the fundamental role as it played in the first tests of General Relativity? In the present paper I argue that it might be the case.

Let us start with the mention of two recent theoretical proposals related to the mystery of dark matter.

In a series of papers (Blanchet 2007a, 2007b; Blanchet and Tiec 2008, 2009) it was suggested that dark matter might be a dipolar fluid composed from gravitational dipoles (in analogy with an electric dipole, a gravitational dipole is defined as a system composed of two particles having the gravitational charge of opposite sign). Hence, Blanchet and Tiec have proposed a dipolar fluid as a new candidate for non-baryonic dark matter; the galaxy rotation curves can be considered as a result of the gravitational polarization of the dipolar fluid by the field of the immersed baryonic matter.

In parallel with the work of Blanchet and Tiec, Hajdukovic (2007, 2008, 2011a, 2011b, 2012a, 2012b) has introduced an apparently similar, but in fact radically different and more economical hypothesis: *quantum vacuum contains virtual gravitational dipoles*. Quantum vacuum is well established in quantum field theories and should be considered as a new state of matter-energy, composed from short living particle-antiparticle pairs (see for instance Aitchison, 2009, for a review). The most elegant possibility is the universal gravitational repulsion between particles and antiparticles, i.e. that all virtual pairs are gravitational dipoles. However, it may also be that the gravitational charge of opposite sign is not a universal property but limited to some particular kind of virtual pairs.

In the following considerations it is only the existence of the gravitational dipoles (and not their nature) that is important. While we consider the case of the gravitational dipoles as inherent part of the quantum vacuum, the qualitative picture should be the same in the case of a dipolar fluid introduced by Blanchet and Tiec.

If the quantum vacuum contains the virtual gravitational dipoles, the gravitational field of a body (Sun, Earth…) immersed in the quantum vacuum, should produce vacuum polarization characterized with a gravitational polarization density $\vec{P}_g$ (i.e. the gravitational dipole moment per unit volume).

As well known, in a dielectric medium, the spatial variation of the electric polarization generates a charge density $\rho_b = -\nabla \cdot \vec{P}$, known as the bound charge density. In an analogous way, the gravitational polarization of the quantum vacuum should result in a gravitational bound charge density

$$\rho_{dm} = -\nabla \cdot \vec{P}_g \quad (1)$$

which might explain the phenomena usually attributed to dark matter.

Assuming the spherical symmetry Eq. (1) reduces to

$$\rho_{dm}(r) = \frac{1}{r^2}\frac{d}{dr}(r^2 P_g(r)) \quad (2)$$

with $P_g(r) \equiv |\vec{P}_g(r)|$.

In principle the space around a spherical body can be divided in two regions. In the inner region, up to a critical distance, the gravitational field is sufficiently strong to align all dipoles along the field; consequently $P_g(r)$ has a constant value (in fact a maximum value) that may be denoted $P_{g\max}$. If $P_{g\max}$ exists it should be considered as a fundamental constant of the quantum vacuum.

Eq. (2) leads to

$$\rho_{dm}(r) = \frac{2P_{g\max}}{r} \quad (3)$$

According to (3) the mass of "dark matter" enclosed within a sphere of radius $r$ is

$$M_{dm}(r) = 4\pi P_{g\max} r^2 \quad (4)$$

and the corresponding acceleration is

$$g_{dm}(r) = -\frac{GM_{dm}(r)}{r^2} = -4\pi G P_{g\max} \equiv A_r \quad (5)$$

The previous studies (Hajdukovic 2011a, 2012a) suggest that the value of $P_{g\max}$ is about one order of magnitude smaller than unity, but the true value must be determined by experiments. On the basis of that result and the new work that is in progress, the best estimate is that $A_r$ lies in the following interval

$$10^{-11} m/s^2 < A_r < 10^{-10} m/s^2 \quad (6)$$

Hence, in the region of saturation (the region of a complete alignment of gravitational dipoles) the Newton force of gravity might be accompanied with a tiny radial force that is a signature of the quantum vacuum.

Before to continue, let us note that, outside the region of saturation, the gravitational polarization density $P_g(r)$ decreases with distance, what, as shown in previous papers (Hajdukovic 2011a, 2012a) has potential to explain the observed phenomena without invoking dark matter. However, we are interested here only in the region of saturation.

Let us forget for the moment the quantum vacuum enriched with the gravitational dipoles and let us turn to a well-established phenomenon: orbits of the planets slowly rotate (precess) in space. Calculations based on the combination of Newtonian mechanics with the inverse square law of gravitation are close to the observed values, with exception of the Mercury and Venus (the planets closest to the Sun). For instance, Mercury's orbit is precessing at a rate that is about 8% greater than that predicted by the Newtonian model. This discrepancy has been explained in the framework of General Relativity by simple formula

$$\Delta\omega_{gr} = \frac{6\pi G}{c^2(1-e^2)}\frac{M_S}{a} \tag{7}$$

where $\Delta\omega_{gr}$ is the extra rotation per orbit in radians, $M_S$ the mass of the Sun, $a$ the semi-major axis of the orbit and $e$ is the eccentricity of the ellipse. Let us note that the general relativistic correction (7) decreases when the semi-major axis increases, what is easy to understand because the general relativistic effects are larger in stronger gravitational fields.

Now, let us return to the quantum vacuum. If relation (5) is correct, the quantum vacuum also contributes to the precession of perihelion.

In the case of two bodies with masses $M$ and $m$, the perihelion precession induced by a constant gravitational acceleration $A_r$ is described with the equation (Murray and Dermott, 1999)

$$\frac{d\omega}{dt} = -\frac{A_r}{e}\sqrt{\frac{a}{M+m}(1-e^2)}\cos f \tag{8}$$

where $f$ denotes the true anomaly.

A simple integration (after expressing $\cos f$ and $dt$ through the eccentric anomaly $E$ gives the shift over the time period $\Delta t$

$$\Delta\omega_{qv} = A_r\sqrt{\frac{1-e^2}{G}}\sqrt{\frac{a}{M+m}}\Delta t; \quad A_r \equiv -4\pi G P_{g\max} \tag{9}$$

In $\Delta\omega_{qv}$ we have used the subscript $qv$ to underline that it is a contribution of the quantum vacuum.

There are two very important points concerning Eq. (9).

First, contrary to the general relativistic contribution (7), the quantum vacuum contribution (9) increases with increase of semi-major axis. Hence, the best natural "laboratories" to study the perihelion precession caused by general relativity and quantum vacuum, must lie in two different limits of the Solar System, near the centre and at the periphery.

Second, the contribution of the quantum vacuum has a negative sign. Hence, if there is a system in which perihelion precession is dominated by the quantum vacuum, we must observe the retrograde precession.

It is obvious that this phenomenon might eventually be revealed by more accurate studies of orbits of planets and orbits of the artificial Earth satellites.

The numerical values $\Delta\omega_{qv}$ for the eight planets of the Solar System are given in Table 1. In these *illustrative* calculations we have used the middle point of the interval (6). It should be noted that these values correspond to the case of perfect spherical symmetry; in general the values are smaller if the spherical symmetry is only approximate. It would be subject of a forthcoming publication but one of the reasons can be easily understood. In the region of saturation, the magnitude of the gravitational polarization density $\vec{P}_g$ is a constant ($P_{g\max}$); if $\vec{P}_g$ deviates from the radial direction, the radial component of $\vec{P}_g$ has a magnitude smaller than $P_{g\max}$ and consequently, the values in Table 1 are upper bounds. For the purpose of comparison Table 1 also contains the observed values (from the book of Fitzpatrick, 2012) and the general relativistic corrections (7).

*Table 1: The observed perihelion precession rates ($\Delta\omega_{obs}$), general relativistic correction ($\Delta\omega_{gr}$) and predicted quantum vacuum contribution ($\Delta\omega_{qv}$) corresponding to $A_r = 5\times 10^{-11} m/s^2$. The precession rates are in arc seconds per century.*

| Planet | $\Delta\omega_{obs}$ | $\Delta\omega_{gr}$ | $\Delta\omega_{qv}$ |
|---|---|---|---|
| Mercury | 575 | 43 | -0.69 |
| Venus | 204 | 8.6 | -0.93 |
| Earth | 1145 | 3.8 | -1.09 |
| Mars | 1628 | 1.4 | -1.34 |
| Jupiter | 655 | 0.06 | -2.48 |
| Saturn | 1950 | 0.01 | -3.36 |
| Uranus | 334 | 0.002 | -4.77 |
| Neptune | 36 | 0.0008 | -5.98 |

According to the previous analysis (Iorio 2006, Page et al. 2009) the current ephemerides of planets do not preclude the illustrative values $\Delta\omega_{qv}$ in the Table 1; and of course it must be kept in mind that these values may be up to five times smaller. Hopefully, the accuracies of the observational data fit by ephemerides would increase about one order of magnitude in the near future, what would allow to confirm or to preclude a perturbation $A_r$ in the interval defined by (6).

As it can be seen from Table 1, for all planets, the effect induced by the quantum vacuum is a small fraction of the observed perihelion precession. It is evident that contribution $\Delta\omega_{qv}$ is small because of the large solar mass in the ratio $a/M$ in the equation (9). Even at a distance of 100AU from the Sun $\Delta\omega_{qv} \approx -10\, arc\sec onds/century$. The small value of $\Delta\omega_{qv}$ is inherent to a Sun-planet system. However, much larger ratio $a/M$ exists in many planet-moon systems. Hence, the ideal "laboratory" to study the precession of the perihelion, induced by the quantum vacuum, is a satellite, orbiting around a dwarf planet on an elliptical trajectory with a large semi-major axis $a$; and of course this system (a dwarf planet with a moon) must be very far from the Sun, in a very weak external gravitational field. In the Solar system, the closest to this idealisation is Eris, a recently discovered dwarf planet (Brown et al. 2005, 2006, and 2007) with a satellite named Dysnomia. The mass of the Eris is $M_{Eris} \approx 1.67\times 10^{22} kg$ and it is presently at a distance of 96 AU from the Sun.

Dysnomia orbits around Eris in an elliptical orbit with small eccentricity $e < 0.013$ and semi-major axis $a = 3.735 \times 10^7 m$.

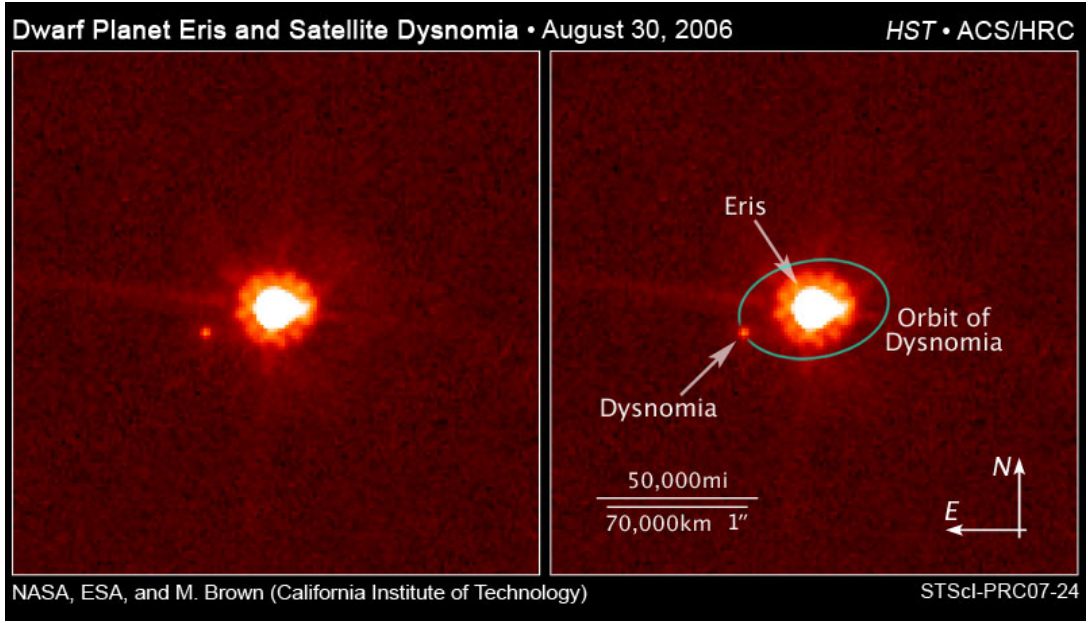

Let us consider Eris and Dysnomia as an ideal two body system. In that case, according to the standard theory there is no perihelion precession, while, according to the relation (9) quantum vacuum induces a precession rate

$$(\Delta \omega_{qv})_{Dy} \approx -190 \, arc\sec/century \qquad (10)$$

Of course, Eris and Dysnomia are not an isolated system; however, the general relativistic correction is miniscule, while the Newtonian contribution can be easily estimated.

The classical (Newtonian) precession rate of a satellite orbit under the solar influence is approximately (see for instance Cuk and Burns, 2004)

$$\left\langle \frac{d\omega}{dt} \right\rangle \approx \frac{3}{4} \frac{n_p^2}{n_s} \equiv \frac{3\pi}{2} \frac{T_s}{T_p^2} \qquad (11)$$

where $n_p$ and $n_s$ are respectively the planet's and the satellite's mean motion, while $T_p$ and $T_s$ are the corresponding periods. For Eris and Dysnomia the periods are 560 years and 16 days. Hence, the classical perihelion precession rate is

$$(\Delta \omega_N)_{Dy} \approx 13 \, arc\sec/century \qquad (12)$$

A Comparison of the numerical results (10) and (12) shows what we have expected; the perihelion precession of Dysnomia (around Eris) is strongly dominated by the quantum vacuum. It is nearly an ideal system for a test of the gravitational properties of the quantum vacuum. A potential difficulty (I am thankful to Professor M.E. Brown for this clarification) is that the orbit of Dysnomia might have significantly smaller eccentricity than the established upper bound $(0.013)$. If so, the measurement of the perihelion precession rate of Dysnomia might become a very challenging task. But, in principle, even if it cannot be done immediately, we should keep in mind that such a possibility exists. The ultimate possibility, allowing much higher precision, would be to study (instead of the orbit of Dysnomia) the orbit of an artificial (human-made) satellite of Eris; of course if the future scientific considerations justify such a complicated and expensive experiment.

Independently of the theoretical speculations, between all objects in the Solar system, Eris-Dysnomia is the closest one to the physical idealisation of an isolated two-body system. In the framework of the Newtonian physics (the general relativistic corrections are many orders of magnitude smaller) the precession of perihelion of Dysnomia has two sources: (a) the external source (i.e. the rest of the Solar System) produces a precession rate roughly estimated by (12). (b) Internal source, mainly a departure from the perfect spherical symmetry of Eris, induces an unknown precession rate, but, as we know from the study of the other cases with approximate spherical symmetry, it is not a dominant part of precession; it is only a fraction of (12). Hence, the eventual discovery of a perihelion precession rate, sharply different from the calculated one in the framework of the Newtonian mechanics, would be a strong sign of new physics.